\documentclass[a4paper,12pt]{article}
\usepackage{amsmath}
\usepackage{amssymb,amsfonts}
\usepackage[english]{babel}
\usepackage{gastex}
\usepackage{graphicx}
\usepackage{cite}
\usepackage[margin=3cm]{geometry}
\begin{document}
\title{Numerical values of the growth rates of power-free languages}
\author{Arseny M. Shur\\Ural State University}
\date{}
\maketitle

\begin{abstract} 
We present upper and two-sided bounds of the exponential growth rate for a wide range of power-free languages.
All bounds are obtained with the use of algorithms previously developed by the author.
\end{abstract}

This is not a research paper but rather an appendix to several research papers by the author, see \cite{Sh08csr,Sh09dlt,ShGo10,Sh10tcs,Sh10csr}. We put together  the best currently known numerical bounds on the growth rate of power-free languages over the alphabets with 2 to 15 letters. Some of these bounds were already published in the mentioned papers, some other are new. All upper bounds are obtained by the algorithm announced in \cite{Sh08csr} and described in full extent in \cite{Sh10tcs}. For all $\beta$-power-free languages with $\beta\ge2$, the obtained upper bounds were converted to the two-sided ones using the results of \cite{Sh09dlt}. 

Recall that the (exponential) growth rate $\alpha(L)$ of a factorial language $L$ is defined by $\alpha(L)=\lim_{n\to\infty}(C_L(n))^{1/n}$, where $C_L(n)$ is the number of words of length $n$ in $L$. We write $\alpha(k,\beta)$ for the growth rate of the $k$-ary $\beta$-power-free language. All upper bounds are obtained as growth rates of some regular languages. Such a ``regular approximation'' of the target power-free language consists of all words avoiding all forbidden powers having the periods bounded from above by some constant $m$. (For example, take the cube-free language over the alphabet $\{a,b\}$ and $m=1$. The corresponding regular approximation is the language of all binary words having no factors $aaa$ and $bbb$. Its growth rate is the golden ratio; this is the simplest nontrivial upper bound for $\alpha(2,3)$.) Lower bounds are obtained from the upper ones by a very simple (almost constant-time) computation.

\medskip
We present a separate table for binary languages; three other tables are devoted to the languages over different alphabets, avoiding ``big'', ``average'', and ``small'' powers, respectively. All bounds are rounded off to 7 decimal places. For $\beta$-power-free languages with $\beta<2$, we give only upper bounds. We also present our estimation of the actual value of the growth rate of each of these languages; such an estimation is obtained by extrapolation from a series of successive upper bounds.

The growth of some particular power-free languages was extensively studied by different authors. The binary $2^+$-power-free language is known to have polynomial growth, so its growth rate equals 1. The same is true for binary $(7/3)$-power-free language. For the growth rate of the binary cube-free language we present the upper bound 1.4575772869240. The estimated exact value of this growth rate is about 1.4575772869237 ($3{\cdot}10^{-13}$ from the upper bound). The best proved lower bound is 1.4575732 ($4{\cdot}10^{-6}$ from the upper bound). Finally, for the growth rate of the ternary square-free language our best upper bound is 1.301761876, the estimated exact value is about 1.30176183, and the best lower bound is 1.3017597.

\medskip
Table \ref{tabbin} shows the behaviour of the growth rate of binary power-free languages. All points in which the function $\alpha(2,\beta)$ jumps by at least 0.001 are included in this table. The sum of jumps in these points is about 0.98. Thus, the behaviour of $\alpha(2,\beta)$ can be seen in details. For $\beta\ge7/2$, we use lower bounds with ``double precision''. The validity of such bounds for the case $\beta\ge4$ was proved in \cite{Sh09dlt}; the argument of \cite{Sh09dlt} can be strenghthened to lower the bound on $\beta$ to $7/2$.

\begin{table}[!htb]
%\vspace*{-1.5mm}
\caption{Binary power-free languages. \small\sl If a cell contains only one number, this number is the exact growth rate of the binary $\beta$-free language, rounded off to 7 decimal places. Otherwise, the cell contains the lower and the upper bounds to such a growth rate; these bounds are also rounded off to 7 decimal places. The integer in the same cell is the number $m$ of the corresponding approximation. The amount of jump of the growth rate at the point $\beta$ is shown in the last column.} \label{tabbin}
\vspace*{4mm}
\tabcolsep=3pt
{\footnotesize\centerline{ 
\begin{tabular}{|r|rc|rc|r|}
\hline
$\beta$&\multicolumn{2}{c|}{$\beta$-free}&\multicolumn{2}{c|}{$\beta^+$-free}&\multicolumn{1}{c|}{jump at $\beta$}\\
\hline
7/3&&1.0000000&65&1.2206318--1.2206448&0.2206\\
\hline
17/7&64&1.2222235--1.2222380&63&1.2287081--1.2287205&0.0065\\
\hline
5/2&62&1.2294871--1.2295017&44&1.3662971--1.3663011&0.1368\\
\hline
18/7&43&1.3669547--1.3669601&43&1.3692782--1.3692832&0.0023\\
\hline
13/5&43&1.3693912--1.3693962&42&1.3760821--1.3760876&0.0067\\
\hline
8/3&42&1.3762649--1.3762704&37&1.4508577--1.4508611&0.0746\\
\hline
14/5&36&1.4522648--1.4522680&36&1.4552314--1.4552358&0.0030\\
\hline
17/6&36&1.4552552--1.4552596&36&1.4567773--1.4567815&0.0015\\
\hline
3&36&1.4575732--1.4575773&24&1.7951246--1.7951264&0.3375\\
\hline
13/4&24&1.7957598--1.7957616&24&1.7972871--1.7972888&0.0015\\
\hline
10/3&24&1.7973088--1.7973105&24&1.8029861--1.8029877&0.0057\\
\hline
7/2&&1.8032409&&1.8172665&0.0140\\
\hline
11/3&&1.8174176&&1.8204960&0.0031\\
\hline
4&&1.8211000&&1.9208015&0.0997\\
\hline
9/2&&1.9214442&&1.9241348&0.0027\\
\hline
5&&1.9244437&&1.9646285&0.0402\\
\hline
6&&1.9653118&&1.9832942&0.0180\\
\hline
7&&1.9834409&&1.9918972&0.0085\\
\hline
8&&1.9919310&&1.9960151&0.0041\\
\hline
9&&1.9960232&&1.9980255&0.0020\\
\hline
\end{tabular} } 
}
\end{table}

Our Table \ref{tabbig} contains the growth rates of the $\beta$-power-free languages with $\beta\ge2$. Some obvious laws of the behaviour of the function $\alpha(k,\beta)$ can be seen in this table; see \cite{Sh10csr} for details.

\begin{table}[!htb]
%\vspace*{-1.5mm}
\caption{Avoiding big exponents: $\beta\ge2$. \small\sl If a cell contains one number, this number is the exact growth rate of the $k$-ary $\beta$-free language, rounded off to 7 decimal places. Two numbers in a cell are the lower and the upper bounds of such a growth rate; these bounds are also rounded off to 7 decimal places.} \label{tabbig}
\vspace*{4mm}
\tabcolsep=3pt
{\footnotesize\centerline{ 
\begin{tabular}{|r|r|r|r|r|r|r|}
\hline
$k\big\backslash\,\beta$&\multicolumn{1}{c|}{$2$}&\multicolumn{1}{c|}{$2^+$}&\multicolumn{1}{c|}{$3$}&\multicolumn{1}{c|}{$3^+$}&\multicolumn{1}{c|}{$4$}&\multicolumn{1}{c|}{$4^+$}\\
\hline
&1.3017597--&2.6058789--&2.7015614--&2.9119240--&&\\
3&1.3017619&2.6058791&2.7015616&2.9119242&2.9172846&2.9737546\\
\cline{2-5}
4&2.6215080&3.7284944&3.7789513&3.9487867&3.9507588&3.9879972\\
5&3.7325386&4.7898507&4.8220672&4.9662411&4.9671478&4.9935251\\
6&4.7914069&5.8277328&5.8503616&5.9760100&5.9764861&5.9961170\\
7&5.8284661&6.8537250&6.8705878&6.9820558&6.9823298&6.9974912\\
8&6.8541173&7.8727609&7.8858522&7.9860649&7.9862337&7.9982866\\
9&7.8729902&8.8873424&8.8978188&8.9888625&8.9889721&8.9987785\\
10&8.8874856&9.8988872&9.9074705&9.9908932&9.9909674&9.9990989\\
11&9.8989813&10.9082635&10.9154294&10.9924142&10.9924662&10.9993163\\
12&10.9083279&11.9160348&11.9221106&11.9935831&11.9936207&11.9994691\\
13&11.9160804&12.9225835&12.9278022&12.9945010&12.9945288&12.9995796\\
14&12.9226167&13.9281788&13.9327109&13.9952350&13.9952560&13.9996615\\
15&13.9282035&14.9330157&14.9369892&14.9958311&14.9958473&14.9997234\\
\hline
\end{tabular} } 
}
\end{table}

\medskip
In Table \ref{tabave} we study the powers that are less than 2 but still relatively big. Since the values $\alpha(k,\frac{3}{2}^+)$ and $\alpha(k,2)$ are quite close to each other for any $k\ge4$, we do not consider the powers $\beta$ such that $\frac{3}{2}^+<\beta<2$. The laws observed in Table \ref{tabave} are also explained in \cite{Sh10csr}.

\begin{table}[!htb]
%\vspace*{-1.5mm}
\caption{Avoiding average exponents: $5/4\le\beta\le(3/2)^+$. \small\sl First row of each cell contains the best obtained upper bound for the growth rate of the $k$-ary $\beta$-free language, rounded off to 7 decimal places, and the number $m$ of the corresponding approximation. The bottom number in the cell is our approximation to the growth rate of the $k$-ary $\beta$-free language obtained by the extrapolation of the series of upper bounds.} \label{tabave}
\vspace*{4mm}
\tabcolsep=2pt
{\footnotesize\centerline{ 
\begin{tabular}{|r|rl|rl|rl|rl|rl|rl|}
\hline
$k\big\backslash\beta$\!&\multicolumn{2}{c|}{$5/4$}&\multicolumn{2}{c|}{$(5/4)^+$}&\multicolumn{2}{c|}{$4/3$}&\multicolumn{2}{c|}{$(4/3)^+$}&\multicolumn{2}{c|}{$3/2$}&\multicolumn{2}{c|}{$(3/2)^+$}\\
\hline
4&&0&&0&&0&&0&\!154&\phantom{1}1.0968025&21&\phantom{1}2.2805723\\
&&&&&&&&&&\sf\phantom{1}1.09679&&\sf\phantom{1}2.28052\\
\hline
5&&0&\!112&\phantom{1}1.1580040&\!109&\phantom{1}1.1646054&24&\phantom{1}2.2489630&22&\phantom{1}2.4024448&16&\phantom{1}3.4928277\\
&&&&\sf\phantom{1}1.1577&&\sf\phantom{1}1.1645(\!6\!)&&\sf\phantom{1}2.2485&&\sf\phantom{1}2.40242(\!4\!)&&\sf\phantom{1}3.492800\\
\hline
6&91&\phantom{1}1.2289301&25&\phantom{1}2.2781859&24&\phantom{1}2.3765882&17&\phantom{1}3.4008105&17&\phantom{1}3.5405292&14&\phantom{1}4.6034033\\
&&\sf\phantom{1}1.2288&&\sf\phantom{1}2.277(\!3\!)&&\sf\phantom{1}2.3763(\!9\!)&&\sf\phantom{1}3.4003(\!9\!)&&\sf\phantom{1}3.540514&&\sf\phantom{1}4.603386\\
\hline
7&26&\phantom{1}2.3267948&19&\phantom{1}3.3737870&19&\phantom{1}3.4660659&15&\phantom{1}4.5026780&15&\phantom{1}4.6275095&13&\phantom{1}5.6727100\\
&&\sf\phantom{1}2.3265&&\sf\phantom{1}3.3729&&\sf\phantom{1}3.4658(\!8\!)&&\sf\phantom{1}4.50224&&\sf\phantom{1}4.627498&&\sf\phantom{1}5.672703\\
\hline
8&20&\phantom{1}3.4104475&16&\phantom{1}4.4532917&16&\phantom{1}4.5400020&14&\phantom{1}5.5749102&14&\phantom{1}5.6867745&13&\phantom{1}6.7206899\\
&&\sf\phantom{1}3.4101&&\sf\phantom{1}4.4515&&\sf\phantom{1}4.53976&&\sf\phantom{1}5.57467&&\sf\phantom{1}5.686769&&\sf\phantom{1}6.720687\\
\hline
9&18&\phantom{1}4.4785124&16&\phantom{1}5.5160143&15&\phantom{1}5.5982898&14&\phantom{1}6.6287238&14&\phantom{1}6.7296776&13&\phantom{1}7.7560582\\
&&\sf\phantom{1}4.4778&&\sf\phantom{1}5.5150(\!2\!)&&\sf\phantom{1}5.59814&&\sf\phantom{1}6.62861&&\sf\phantom{1}6.729676&&\sf\phantom{1}7.756057\\
\hline
10&17&\phantom{1}5.5342518&15&\phantom{1}6.5670585&15&\phantom{1}6.6443657&14&\phantom{1}7.6703573&14&\phantom{1}7.7621761&12&\phantom{1}8.7832930\\
&&\sf\phantom{1}5.5336(\!4\!)&&\sf\phantom{1}6.5663&&\sf\phantom{1}6.64428&&\sf\phantom{1}7.67029(\!4\!)&&\sf\phantom{1}7.762175&&\sf\phantom{1}8.783291\\
\hline
11&17&\phantom{1}6.5802111&15&\phantom{1}7.6086719&15&\phantom{1}7.6813567&14&\phantom{1}8.7035224&14&\phantom{1}8.7876552&12&\phantom{1}9.8049486\\
&&\sf\phantom{1}6.57983&&\sf\phantom{1}7.6081(\!7\!)&&\sf\phantom{1}7.68131&&\sf\phantom{1}8.70348(\!6\!)&&\sf\phantom{1}8.787655&&\sf\phantom{1}9.804948\\
\hline
12&17&\phantom{1}7.6185200&15&\phantom{1}8.6432160&15&\phantom{1}8.7115615&14&\phantom{1}9.7305688&13&\phantom{1}9.8081756&12&10.8226038\\
&&\sf\phantom{1}7.61827&&\sf\phantom{1}8.6428(\!8\!)&&\sf\phantom{1}8.71153(\!5\!)&&\sf\phantom{1}9.730547&&\sf\phantom{1}9.808175&&\sf10.822603\\
\hline
13&17&\phantom{1}8.6507655&15&\phantom{1}9.6722706&15&\phantom{1}9.7366286&14&10.7530530&13&10.8250601&12&11.8372861\\
&&\sf\phantom{1}8.65058(\!9\!)&&\sf\phantom{1}9.67204&&\sf\phantom{1}9.736612&&\sf10.753039&&\sf10.825060&&\sf11.837286\\
\hline
14&17&\phantom{1}9.6781738&15&10.6970094&15&10.7577374&14&11.7720444&13&11.8392005&12&12.8496956\\
&&\sf\phantom{1}9.67804(\!8\!)&&\sf10.69685&&\sf10.757727&&\sf11.772035&&\sf11.839200&&\sf12.849695\\
\hline
15&17&10.7017089&15&11.7183082&15&11.7757426&14&12.7883022&13&12.8512175&12&13.8603271\\
&&\sf10.701616&&\sf11.71819&&\sf11.775735&&\sf12.788296&&\sf12.851217&&\sf13.860327\\
\hline
\end{tabular} } 
}
\end{table}

\medskip
The results given in Table \ref{tabsma} concern about small power-free languages. It is known from Dejean's conjecture (finally proved in 2009) that the $k$-ary $\beta$-power-free language is infinite iff $\beta\ge(7/4)^+$ for $k=3$, $\beta\ge(7/5)^+$ for $k=4$, and $\beta\ge{\frac{k}{k{-}1}\!}^+$ for all other $k$. To make Table \ref{tabsma} readable, we give the information about the minimal power-free languages over the ternary and quaternary alphabets in the column headed by ${\frac{k}{k{-}1}\!}^+$. The conjectures about the behaviour of the function $\alpha(k,\beta)$ for the case of small $\beta$ are given in \cite{ShGo10,Sh10csr}.

\begin{table}[!htb]
%\vspace*{-1.5mm}
\caption{Avoiding small exponents: $\beta\le(k{-}3)/(k{-}4)$. \small\sl First row of each cell contains the best obtained upper bound for the growth rate of the $k$-ary $\beta$-free language, rounded off to 7 decimal places, and the number $m$ of the corresponding approximation. The bottom number in the cell is our approximation to the growth rate of the $k$-ary $\beta$-free language obtained by the extrapolation of the series of upper bounds.} \label{tabsma}
\vspace*{4mm}
\tabcolsep=2pt
{\footnotesize\centerline{ 
\begin{tabular}{|r|rl|rl|rl|rl|rl|rl|}
\hline
$k\big\backslash\beta$\!&\multicolumn{2}{c|}{${\dfrac{k}{k{-}1}\!}^+$}&\multicolumn{2}{c|}{$\dfrac{k{-}1}{k{-}2}$}&\multicolumn{2}{c|}{$\dfrac{k{-}1}{k{-}2}^+$}&\multicolumn{2}{c|}{$\dfrac{k{-}2}{k{-}3}$}&\multicolumn{2}{c|}{$\dfrac{k{-}2}{k{-}3}^+$}&\multicolumn{2}{c|}{$\dfrac{k{-}3}{k{-}4}$}\\
\hline
3*&66&1.2456093&&&&&&&&&&\\
&&\sf1.245608&&&&&&&&&&\\
\hline
4*&208&1.0695061&\!156&1.0968016&21&2.2805723&&&&&&\\
&&\sf1.0694&&\sf1.09679&&\sf2.28052&&&&&&\\
\hline
5&\!115&1.1579787&\!112&1.1645978&24&2.2489630&22&2.4024448&16&3.4928277&&\\
&&\sf1.1577&&\sf1.1645(\!6\!)&&\sf2.2485&&\sf2.40242(\!4\!)&&\sf3.492800&&\\
\hline
6&95&1.2247121&92&1.2289256&25&2.2781859&24&2.3765882&17&3.4008105&17&3.5405292\\
&&\sf1.2246&&\sf1.2288&&\sf2.277(\!3\!)&&\sf2.3763(\!9\!)&&\sf3.4003(\!9\!)&&\sf3.540514\\
\hline
7&100&1.2369024&99&1.2373991&27&2.299738&26&2.3267948&19&3.3737870&19&3.4660659\\
&&\sf1.2368(\!6\!)&&\sf1.2373(\!7\!)&&\sf2.2990&&\sf2.3265(\!7\!)&&\sf3.3729&&\sf3.4658(\!8\!)\\
\hline
8&109&1.2348427&100&1.2349430&29&2.3132472&29&2.3285295&20&3.3639604&20&3.4104475\\
&&\sf1.23483&&\sf 1.23494&&\sf2.312(\!7\!)&&\sf2.327(\!0\!)&&\sf3.361(\!7\!)&&\sf3.4101\\
\hline
9&107&1.2466776&100&1.2467443&31&2.3194735&31&2.3261560&22&3.3639868&22&3.3904233\\
&&\sf1.24667&&\sf 1.24674&&\sf2.319(\!0\!)&&\sf2.326(\!0\!)&&\sf3.362&&\sf3.390\\
\hline
10&109&1.2393075&100&1.2393375&31&2.3235082&32&2.3259292&23&3.3669859&23&3.3796610\\
&&\sf1.239307&&\sf 1.239337&&\sf2.322(\!9\!)&&\sf2.325(\!5\!)&&\sf3.364&&\sf3.377\\
\hline
11&110&1.2426060&92&1.2426389&31&2.3273701&31&2.3273701&25&3.3693164&25&3.3759580\\
&&\sf1.242606&&\sf 1.242638(\!7\!)&&\sf2.324&&\sf2.325&&\sf3.367&&\sf3.374\\
\hline
12&108&1.2428777&96&1.2428801&33&2.3289650&33&2.3289650&26&3.3714325&27&3.3745058\\
&&\sf1.242877(\!5\!)&&\sf1.242880&&\sf2.325&&\sf2.326&&\sf3.370&&\sf3.374\\
\hline
13&104&1.2408703&96&1.2408753&36&2.3294446&36&2.3294446&26&3.3760035&26&3.3762406\\
&&\sf1.240870(\!1\!)&&\sf1.240875&&\sf2.326&&\sf2.326&&\sf3.372&&\sf3.37(\!5\!)\\
\hline
14&105&1.2427746&98&1.2427761&38&2.3297907&38&2.3297907&26&3.3863857&26&3.3864742\\
&&\sf1.242774(\!2\!)&&\sf1.242776&&\sf2.326&&\sf2.326&&\sf3.374&&\sf3.37(\!6\!)\\
\hline
15&105&1.2418324&90&1.2418340&41&2.3299169&41&2.3299169&27&3.3942863&27&3.3943189\\
&&\sf1.241832&&\sf1.241833&&\sf2.327&&\sf2.327&&\sf3.375&&\sf3.37(\!7\!)\\
\hline
\end{tabular} } 
}
\end{table}

\end{document}